\documentclass{article}
\usepackage{PRIMEarxiv}

\usepackage[utf8]{inputenc} % allow utf-8 input
\usepackage[T1]{fontenc}    % use 8-bit T1 fonts
\usepackage{hyperref}       % hyperlinks
\usepackage{url}            % simple URL typesetting
\usepackage{booktabs}       % professional-quality tables
\usepackage{amsfonts}       % blackboard math symbols
\usepackage{nicefrac}       % compact symbols for 1/2, etc.
\usepackage{microtype}      % microtypography
\usepackage{lipsum}
\usepackage{fancyhdr}       % header
\usepackage{graphicx} 
\usepackage{svg}
\usepackage{diagbox}
\usepackage{booktabs}
\usepackage{soul}
\usepackage{lipsum}
\usepackage{color}% graphics
\graphicspath{{media/}}    

\pdfoutput=1
\title{Effects of Context, Complexity, and Clustering on Evaluation for Math Formula Retrieval}

\author{
  Behrooz Mansouri \\
  Rochester Institute of Technology \\
  Rochester, NY, USA\\
  \texttt{bm2203@rit.edu} \\
   \And
  Douglas W. Oard \\
  University of Maryland \\
  College Park, USA\\
  \texttt{oard@umd.edu} \\
     \And
  Anurag Agarwal \\
  Rochester Institute of Technology \\
  Rochester, NY, USA\\
  \texttt{axasma@rit.edu} \\
     \And
  Richard Zanibbi \\
  Rochester Institute of Technology \\
  Rochester, NY, USA\\
  \texttt{rxzvcs@rit.edu} \\
}

\begin{document}
\maketitle

\begin{abstract}
There are now several test collections for the formula retrieval task, in which a system's goal is to identify useful mathematical formulae to show in response to a query posed as a formula. 
These test collections differ in query format, query complexity, number of queries, content source, and relevance definition. Comparisons among six formula retrieval test collections illustrate that defining relevance based on query and/or document context can be consequential, that system results vary markedly with formula complexity, and that judging relevance after clustering formulas with identical symbol layouts (i.e., Symbol Layout Trees) can affect system preference ordering.
\keywords{Math Formula Search, Evaluation, Test Collections}
\end{abstract}

\section{Introduction}
To express mathematical information needs, users might use a mathematical formula as their query \cite{mansouri2019characterizing}.
For example, a user could search using the formula for the $F_1$ evaluation measure, and see how similar formulae are used in different domains. As another example, one can search with a formula to retrieve the concept the formula represents. For instance, a user can search $a^2+b^2=c^2$, to find that this is known as the ``Pythagorean Theorem.'' This application is used in search engines such as MathDeck \cite{diaz2021mathdeck}. Finally, formula search can be used when users want to simply solve for the value of a variable in an equation. For example, a user can search using $x^2-10x+25=0$ to find the result $x=5$.

Evaluation of formula retrieval systems is challenging; several issues need to be considered, including formula query selection, technical complexity of the collection, relevance definitions, and evaluation protocols. Formula retrieval has been studied in several shared tasks, including tasks at NTCIR 10, 11 and 12 \cite{ntcir10, aizawa2014ntcir, zanibbi2016ntcir}. More recently, the ARQMath labs at CLEF 2020 and 2021 have had a formula retrieval task \cite{zanibbi2020overview,mansouri2021overview}.
To date, most formula retrieval tasks have been ad hoc search tasks, the exception being NTCIR-11, which was a known-item retrieval task (i.e., a specific instance of a specific formula was sought). 

From these tasks, a total of six test collections have been created, which we describe and compare in this paper. In particular,
system rankings can differ when using NTCIR and ARQMath collections. This may be due to several differences between the newer ARQMath test collections and the earlier NTCIR test collections. Among the most important differences are: 

\begin{enumerate}
    \item \textbf{Queries.} ARQMath collections have more queries, which are also more diverse (in terms of what we call complexity).
    \item \textbf{Relevance.} In ARQMath, the definition of relevance is more strongly tied to the contexts in which the which formulae originally appeared.
    \item \textbf{Pooling.} In NTCIR, many instances of the same formula might be retrieved, so assessment pools could be rich in duplicates.  In ARQMath, identical formulae were clustered prior to pooling, leading to more diverse pools.
    \item \textbf{Evaluation.} In NTCIR, systems were given credit for every retrieved formula instance that was relevant.  In ARQMath, systems received credit only for the first instance of each  distinct relevant formula retrieved. 
\end{enumerate}

In this paper we describe available formula retrieval test collections, and then share further analysis illustrating that differences in the number of topics and their diversity, the relevance definition, and evaluation protocols can produce different rankings among the same set of formula retrieval systems, and thus some choices made during test collection design and use are consequential. 

\section{Selection of Content Sources and Queries}

Formula retrieval collections have drawn on different sources. NTCIR-10 relied on 100,000 technical papers from arXiv, including papers on mathematics, physics, and computer science. Due to the high technical complexity of arXiv papers, NTCIR-11 and 12 instead made use of Wikipedia articles. 
ARQMath, which is now entering its third year, has two tasks: (1) Answer retrieval and (2) Formula retrieval. In the first task, the goal is to find relevant answers to mathematical questions. In the second task, a formula is chosen from questions in task 1, and the goal is to find relevant formulae. In this paper, we focus only on this formula retrieval task. 
ARQMath uses questions and answers from the Community Question Answering (CQA) website Math Stack Exchange (MSE).\footnote{https://math.stackexchange.com/} In terms of technical complexity, MSE varies from simple questions to expert-level mathematical inquiries. The ARQMath collection contains posts from 2010 to 2018, with queries chosen from 2019 (for ARQMath-1) or 2020 (for ARQMath-2).

\begin{figure*}[!tp] 
\centering
\includegraphics[width=0.8\textwidth]{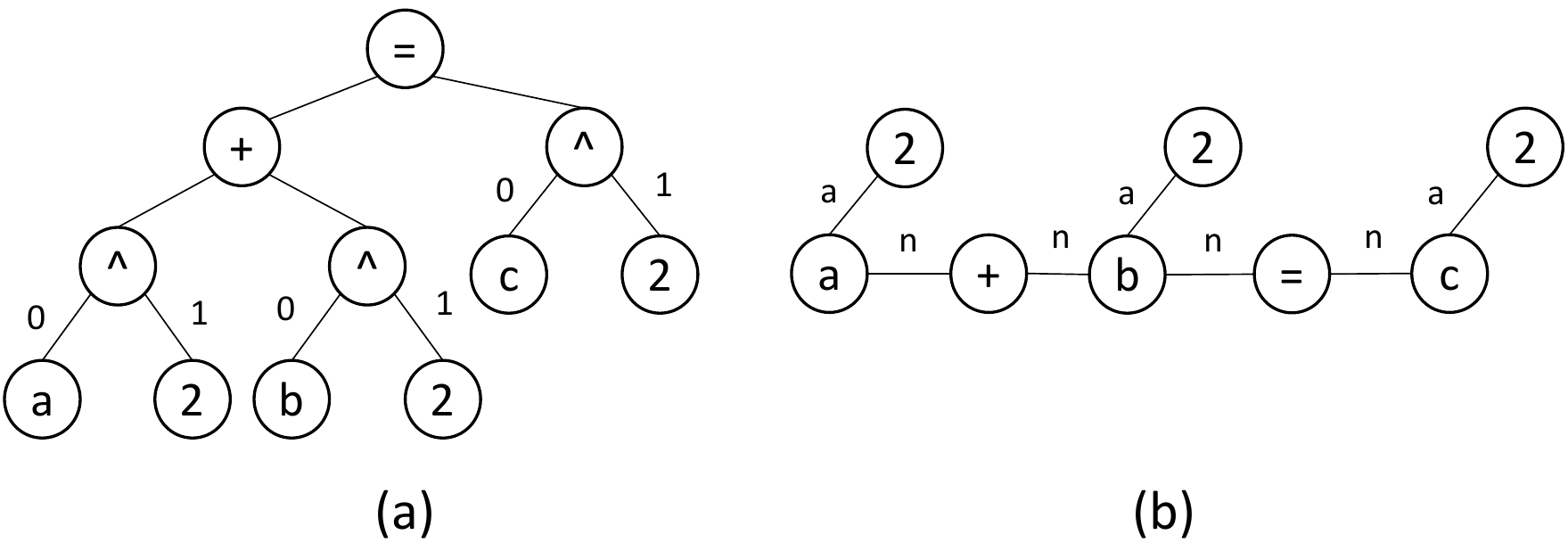}
\caption{Formula \( a^2 + b^2 = c^2 \) as an (a) Operator Tree and (b) Symbol Layout Tree.}
\label{fig:slt_opt}
\end{figure*}

In the NTCIR and ARQMath collections, each formula is identified with a ``math-container'' tag that includes a unique id for each formula instance.  The ARQMath collections and NTCIR-12 collections provide three formula representations: (1) the original \LaTeX~string representation, 
(2) a Symbol Layout Tree (SLT), representing symbol placement on writing lines (similar to \LaTeX), and (3) an Operator tree (OPT), a mathematical syntax encoding giving a hierarchy of operators and their arguments. Figure \ref{fig:slt_opt}
provides schematic examples of OPT and SLT representations.  More specifically, SLTs and OPTs are represented using Presentation MathML and Content MathML markup, respectively.

While formula search systems such as Approach0 \cite{zhong2019structural} use just one type of representation, other systems such as Tangent-S \cite{davila2017layout} use both SLTs and OPTs. Systems such as XY-PHOC \cite{avenoso2021xy} that use neither SLT nor OPT representations can use \LaTeX~as a basis for generating their desired input representation(s). 

{\bf Queries.}
Formula queries in the test collections differ in their number, diversity, and nature. \emph{Wildcard} queries contain symbols that may be replaced by variables or sub-expressions; we refer to other other queries as \emph{concrete}. 

In NTCIR-10, 21 formula queries were chosen by the organizers from arXiv papers. 18 of the 21 queries include wildcards; 3 are concrete. A search scenario and query-specific judgment criteria were specified for each formula query.

In NTCIR-11, a total of 100 queries were selected randomly from Wikipedia pages. 59 queries include wildcards; 41 are concrete. NTCIR-11 was a known-item retrieval task, so no relevance judgment criteria were specified.

In the NTCIR-12 Wikipedia Formula Browsing (WFB) task there were 40 queries, divided into 20 concrete queries and 20 wildcard queries.
The wildcard queries were created by replacing one or more subexpressions in each concrete formula query with wildcards. For example, the query $O(mn\log m)$ also appears as $O(${\bf *1*} $\log$ {\bf *2*}$)$, where {\bf *1*} and {\bf *2*} represent two  subexpressions that may differ. In contrast,  $O(${\bf *1*} $\log$ {\bf *1*}$)$ require the wildcards to match identical subexpressions (e.g., $O({\bf n} \log {\bf n})$).

NTCIR-12 had a second formula retrieval task called \textit{simto}, with 8 queries.\footnote{For conciseness, we refer to the NTCIR-12 WFB task simply as NTCIR-12 except when comparing to the \textit{simto} task.} The \textit{simto} operator identifies subexpressions to be matched by similarity, with other subexpressions matched exactly. 

In ARQMath, a total of 144 concrete formula queries were selected from Math Stack Exchange question posts from a time period that was disjoint from that of the posts to be searched.  In ARQMath-1 there were 74 queries of which 45 were assessed initially and used to evaluate systems (the Test queries), and the remaining 29 being assessed later and thus designated for use in training future systems (the Train queries). In ARQMath-2, a total of 70 formula queries were assessed. 58 queries were used as Test queries and 12 were assessed for the Train set. The Math Stack Exchange question post was used in place of the scenarios that had been provided with NTCIR-10 and NTCIR-12 queries; no other query-specific relevance judgment criteria were provided.

\begin{table}[t!]
\caption{\label{tab:complexities} Concrete query complexity for NTCIR-10/12 and ARQMath-1 and -2.}
\small
\begin{center}
\begin{tabular}{l  l | c  c  | c  c| c  c}
\toprule
 \multicolumn{2}{l|}{\sc Complexity} & \multicolumn{2}{c|}{\sc   NTCIR } & \multicolumn{2}{c}{\sc ~~ ARQMath-1 ~~}  & 
 \multicolumn{2}{c}{\sc ~~ ARQMath-2}  \\ 
 \sc Level & \sc Example  & ~10 & 12 & \sc test &  \sc train& \sc test &  \sc train\\
\midrule
\textbf{Low} & $O(mn\log m)$ & 11  & 4 & 21 & 15&22&4\\
\textbf{Medium~~} & $\int_{0}^{\infty}\frac{\sin x}{x^{a}}$ & 6 & 6 & 16 & 11&24&6\\ 
\textbf{High} & $\sum_{r=1}^n (-1)^{(n-r)}~ {n \choose r}(r)^m$ & 5 & 10 & 8 & 3&12&2\\
\bottomrule
\end{tabular}
\end{center}
\end{table}

ARQMath provided a complexity label for its formula queries. Based on the number and diversity of symbols and structures in a formula, 3 levels of complexity were defined: Low (L), Medium (M), and High (H). For this paper we have manually annotated the NTCIR-10 queries and the 20 concrete NTCIR-12 queries in the same way; these new annotations were checked by the same mathematician who checked the complexity labels for the ARQMath test collection. Table \ref{tab:complexities} shows the distribution of topic complexity for each test collection, along with an example of each complexity level. In NTCIR-10, ARQMath-1 and ARQMath-2, there are more low-complexity queries than in NTCIR-12.

As Mansouri et al.~\cite{mansouri2019characterizing} found, low-complexity formulae are common in queries posed to general-purpose search engines. NTCIR-10 and ARQMath-1/-2 queries provide representative models for that use case. On the other hand, high-complexity queries can be easily produced using cut-and-paste, and there is related work on helping users to find, create, and edit large formulae in search interfaces \cite{diaz2021mathdeck}. So it is also important for test collections to include some high-complexity queries.

\section{Relevance Judgments}

When developing a formula retrieval test collection, two fundamental decisions must be made: (1) which formulae to judge for relevance, and (2) how degrees of relevance should be decided and encoded.  

{\bf Pooling.}
The set of formula instances to be judged was created by pooling submitted runs for all of the test collections except NTCIR-11, the known-item retrieval task. Table~\ref{tab:details} shows the number of runs that were pooled in each case.

\begin{table}[!t]
\caption{\label{tab:details} Formula retrieval collections: sources, number of participating teams, and number of runs.}
\small
\begin{center}
\begin{tabular}{l l  r   r }
\toprule
 \sc Task & \sc Collection & \sc Teams & \sc ~~Runs \\
\midrule
\sc ARQMath-2 & Math Stack Exchange~~  & 6& 18\\
\sc ARQMath-1 & Math Stack Exchange & 3& 11\\
\midrule
\sc NTCIR-12 WFB ~~ & Wikipedia & 2& 7\\
\sc NTCIR-12 \textit{simto} & arXiv & 2& 8\\
\sc NTCIR-11 & Wikipedia & 7& 21\\
\sc NTCIR-10 & arXiv & 6& 12\\
\bottomrule
\end{tabular}
\end{center}
\end{table}

In NTCIR-10 participating teams could submit up to 100 formula instances\footnote{Two formula instances were defined in NTCIR as distinct if they occur in different documents, even if they are otherwise identical.} per formula query.  The highest-ranked formula instances were selected from each submitted run, one rank at a time, until the pool to be judged contained at least 100 unique formula instances, each of which had been highly ranked by at least one system.   In the NTCIR-12 WFB and NTCIR-12 \textit{simto} tasks, participating teams could submit up to 1,000 results per topic.  In each case (separately), the top-20 formula instances from each submitted run were pooled.  

This instance-based definition of uniqueness sometimes results in limited diversity in the judgment pools.  As the most extreme example, for the NTCIR-12 WFB query $\beta$ (a short formula consisting of a single symbol), every formula instance in the judgement pool was $\beta$. To increase the diversity in the judgment pools, ARQMath used an appearance-based definition of uniqueness. 

Specifically, when two formulae have identical symbolic representations they are defined as \emph{visually identical}, and otherwise as \emph{visually distinct}. Symbolic identity was defined as identical Symbol Layout Trees as represented by Tangent-S~\cite{davila2017layout} when both were parseable, and by identical \LaTeX~ strings otherwise. Table \ref{tab:visual_formula} illustrates this distinction using examples of visually identical formulae with different \LaTeX~ representations.

\begin{table}[t!]
\caption{\label{tab:visual_formula} Visually identical formulae with different  \LaTeX{} representations.}
\begin{center}
\begin{tabular}{l |  l  l l}
\toprule
 \multicolumn{1}{l|}{Formula} & \multicolumn{3}{c}{Three Different \LaTeX{} Strings}\\
\midrule
 $a^2=2b^2$ ~~&  \verb|{a^{2}=2b^{2}}| ~~~~& \verb|{a^2}=2{b^2}| ~~~~& \verb|a^2=2b^{2}| ~~~~\\
 $m\neq0$  &  \verb|{m \ne 0}|& \verb|m \not= 0|& \verb|m \ne \\0|\\
 $\frac{n}{m}$ & \verb|{\frac{n}{m}}|& \verb|n \over m|& \verb|\frac n m|\\
\bottomrule
\end{tabular}
\end{center}
\end{table}

At least one run from each team, plus an organizer-provided baseline, was pooled to the depth of the first instance of the 25th (in ARQMath-1) and 20th (in ARQMath-2) visually-distinct formula for that run. For other runs, the pool depth was the first instance of the 10th visually distinct formula.  As Table \ref{tab:annotated} shows, this process yielded the largest judgment pools amongst current formula retrieval test collections. In ARQMath-1, participating teams were asked to return formula instances. For ARQMath-2, each visually-distinct formulae had a predefined unique visual-id, created by clustering the entire collection. Teams had the visual-ids before submitting their runs, so that they could choose which duplicates to submit. 

For each visually distinct formula, at most 5 instances were assessed. This limit was rarely reached: of the 5,843 assessed visually distinct ARQMath-1 formulae, just 93 (1.6\%) had instances in more than 5 pooled posts; for ARQMath-2 the corresponding statistic was 1.4\% (117 of 8,129). When selection was necessary, the 5 instances were selected to favor those submitted in the largest number of runs in ARQMath-1. In ARQMath-2, reciprocal rank fusion was used to give higher weights to instances returned at higher ranks in multiple runs. 

{\bf Judging and Encoding Relevance.} 
Determining the relevance of a retrieved formula to a formula query can be challenging. First, relevance naturally depends on the user's reason for issuing that query, only parts of which may be signaled by the content or form of the query. Second, even with an understanding of that context, mathematical knowledge may be needed to recognize whether a retrieved formula would likely be useful. 

NTCIR-11 was a known-item retrieval task. For each topic, the single Relevant (R) formula instance was defined as the formula instance that had been used as the formula query. All other formula instances, including similar or even identical formulae in the collection would be scored as Non-relevant (N).

For NTCIR-10, the assessors were mathematicians or math students.  Assessors viewed each formula instance from the judgment pool in isolation and assigned each a grade of Relevant (R), Partially relevant (P), or Non-relevant (N) to that pool's query, considering the query-specific scenario and judgment criteria that had been specified when the query was created. 

For the NTCIR-12 WFB task there were two groups of assessors; each group independently judged each pooled formula. One group consisted of computer science graduate students, the other consisted of computer science undergraduates. Assessors viewed each formula instance from the judgment pool for a query in context, and assigned each a relevance grade of Relevant (R), Partially relevant (P), or Non-relevant (N) to that pool's query, informed by the scenario and judgment criteria that had been specified when the query was created. The pooled formula instances were shown to the assessors in context (i.e., highlighted in the text where it had appeared in the collection), but assessors were not asked to interpret the pooled formula in that specific context; the assessment was to be done based on the pooled formula itself, with reference to the scenario and relevance criteria that were provided to the assessor with that query. For the NTCIR-12 \textit{simto} task, two assessors judged each pooled formula, following the same approach as for the NTCIR-12 WFB task.

The ARQMath assessors were undergraduate and graduate mathematics students. Relevance judgments for each pooled formula instance was performed in context, using both the context of the formula query and the context of the pooled formula. The assessor was shown the Math Stack Exchange posts in which the query and the retrieved formula instances appeared, with each formula highlighted,\footnote{Assessors could optionally also examine the question/answer/comment threads in which the question post and retrieved formula instance post had occurred.} and they were asked to judge the extent to which the pooled formula would be useful to the searcher who had issued the formula query. Four relevance grades were defined: Relevant (R, defined as just as good as finding an exact match), Partially relevant high (P+, defined as helpful but not to the same degree as the exact formula), Partially relevant low (P-, defined as even less helpful for finding something useful), and Non-relevant (N).\footnote{We have harmonized the naming of relevance levels across all collections for this paper.  Some of the actual collections used different nomenclature for these categories.} 

\begin{table}[t!]
\caption{\label{tab:definition_example} Example relevance ratings for query $\int _{x=0}^{\infty} \frac{\sin(x)}{x}$ in ARQMath-1.}
\scriptsize
\centering
 \resizebox{\columnwidth}{!}{
\begin{tabular}{l  l   p{3in}  }
\toprule
\sc Rating~~ & \sc Example~~~~~~ & \sc Assessor's Comment \\
\midrule
\sc {R} & $\int_{-\infty}^{\infty}\frac{\sin x}{x}$ & The query is an even function, this is just twice as the query.\\
~\\
\sc{P+} & $\int_0^\infty \frac{\sin(ax)}{x}$ & For arbitrary `a', a change of variable gives same integration.\\
~\\
\sc{P-} & $\sum_{x = 1}^{\infty} \frac{\sin(x)}{x}$ & One tangential possibility is to use the definition of Riemann sum to calculate  the integration.\\
~\\
\sc{N} & $\int_0^{\infty} \sin f(x)$ & Too generalized to be relevant.\\
\bottomrule
\end{tabular}
}
\end{table}

Table \ref{tab:definition_example} shows four formulae with different relevance degrees for the ARQMath-1 query $\int _{x=0}^{\infty} \frac{\sin(x)}{x}$. In this example, the assessor has differentiated relevance degrees based on the generalization applied to the query formula.

The definitions of relevance levels in ARQMath-1 and ARQMath-2 were ostensibly the same, but the basis for the judgments was different.  In both ARQMath-1 and ARQMath-2, the searcher's intent was inferred from the question in which the query had appeared; it was the role of context for retrieved formulas that differed.  ARQMath-1 assessors considered a retrieved formula in isolation, and determined whether a retrieved formula in the pool could productively have been substituted for the query formula; would such a substitution likely find useful results?  As a consequence of this definition, ARQMath-1 assessors were told an exact match would certainly be relevant. However, in ARQMath-2, relevance was decided based on the likelihood of retrieved formulas being associated with information helpful in answering the question, but taking variable and operation types, variable constraints, etc. described in a retrieved formula's post into consideration. As we illustrate later, this means that formulas visually identical to the query may be non-relevant.

\begin{table}[tb!]
\scriptsize
\caption{\label{tab:annotated} Judgements per collection. Here, Relevant includes all but `Non-Relevant'.  
}
{

\begin{center}
\begin{tabular}{l  c  r  r  r  r  r  r }
\toprule

&   & \multicolumn{3}{c}{\sc Judged} & \multicolumn{3}{c}{\sc Relevant} \\
 \sc Collection & \sc Queries & \multicolumn{1}{c}{\sc Min} & \multicolumn{1}{c}{\sc Max} & \multicolumn{1}{c }{\sc Avg } & \multicolumn{1}{c}{\sc Min} & \multicolumn{1}{c}{\sc Max} & \multicolumn{1}{c}{\sc Avg}\\
\midrule
 \multicolumn{8}{c}{ \bf Formula Instances}\\
 \hline
 {NTCIR-10} & 21 & 100  & 105 & 101.3 & 2 & 76 & ~~35.6\\
\sc {NTCIR-11} & 100 & -  & - & - & 1 & 1 & 1.0\\
 {NTCIR-12 \textit{simto}} & 8 & 55 & 113  & 81.8 & 1 &  39& 19.8\\
\hline
 {NTCIR-12 WFB} &  &  &   &  &  & &  \\
 {~~~~Concrete} & 20 & 30 & 90  & 59.5 & 14 & 89 & 41.1 \\
 {~~~~Wildcard} & 20 & 48 & 107  & 74.3 &  12& 80 & 50.9 \\
\hline
 {ARQMath-1}  &  &  &  &  & &  &  \\
 {~~~~Test} & 45 & 97 & ~~206 & ~~147.9 & 6 & 81 & 31.0\\
 {~~~~Train} & 29 & 133 & 198 & 160.1 & 2 & 81 & 29.1\\
\hline
 {ARQMath-2}  &  &  &  &  & &  &  \\
 {~~~~Test} &  58& 110 & 300 & 178.9 & 7 & 175 & 74.0 \\
 {~~~~Train} & 12 & 143 & 214 & 177.7 & 10 & 165 & 87.6 \\
\midrule
 \multicolumn{8}{c}{ \bf Visually Distinct Formulae}\\
\hline
 {NTCIR-12 WFB} &    &  &  &  &  &  & \\
 {~~~~Concrete} & 20 & 1 & 86 & 49.8 & 1 & 63 & 32.7\\
 {~~~~Wildcard} & 20 & 25 & 87 & 61.9 & 11 & 77 & 41.3\\
\hline
 {ARQMath-1}  &  &  &  &  & &  &  \\
 {~~~~Test}  & 45 & 102 & 159 & 134.9 & 6 & 62 & 27.4 \\
 {~~~~Train}  & 29 & 72 & 164 & 134.8 & 2 & 78 & 24.0 \\
\hline
 {ARQMath-2}  &  &  &  &  & &  &  \\
 {~~~~Test} & 58 &  98& 189 & 139.8 & 6 & 115 & 53.4\\
 {~~~~Train} & 12 & 123 & 180 & 144.5 & 6 & 104 & 62.1  \\
\bottomrule
\end{tabular}
\end{center}
}
\end{table}

Table \ref{tab:annotated} summarizes the number of annotated formula instances in each test collection. On average, the ARQMath collections have the largest number of judged formula instances per query.  We have the judgments of individual assessors only for NTCIR-12 and ARQMath. For the NTCIR-12 WFB task, kappa scores for chance-corrected annotator agreement were 0.28; for NTCIR-12 \textit{simto}, kappa was 0.55; for ARQMath-1, kappa was 0.48; for ARQMath-2 kappa was 0.69. There is thus some evidence that the ARQMath-2 relevance definition may have improved external validity, although we note that there were also differences in the assessor training processes between ARQMath-1 and ARQMath-2. 

\section{Evaluation Protocols}
Let us now examine how systems were compared in the evaluation venues for which these test collections were created. We note that the evaluation protocols are separable from the collections: any evaluation protocol could in principle be used with any test collection.

NTCIR-10, NTCIR-12 WFB, and NTCIR-12 \textit{simto} combined 3-level judgments from two assessors to form a 5-level ``Aggregate'' relevance judgment for each assessed document. This was done by mapping N to 0, P to 1 and R to 2, and then summing the two scores.  The resulting integer scores ranged from a low of 0 (both assessors judged N) to a high of 4 (both assessors judged R).  For all three evaluations, evaluation measures were computed on formula instances.  NTCIR-10 reported P@5, P@10, P@hit (i.e., for all returned results), and MAP. NTCIR-12 WFB and NTCIR-12 \textit{simto} reported P@5, P@10, P@15 and P@20. Relevance judgments for formula instances that were missing from the pools (as can happen for P@hit and MAP) were treated as not relevant.  For computing all of these measures, results were binarized by treating aggregated relevance grades 0, 1, and 2 as Non-relevant and treating 3 and 4 as Relevant.

The evaluation measure for the NTCIR-11 known-item retrieval task was the Mean Reciprocal Rank (MRR) of the single relevant formula instance per query.

Recall that in ARQMath, between 1 and 5 instances for each visually distinct formula had been judged for relevance.  To form a relevance judgment for each visually distinct formula, the maximum (i.e., most relevant) judgment (in this preference order: R, P+, P-, N) was used.  For example, if three instances of the same visually distinct formula were judged as N, N and P-, then the aggregate judgment for that visually distinct formula would be P-.

The ARQMath evaluation measures (variants of nDCG, MAP and P@10) were then computed on visually distinct formulae.  Result sets in ARQMath were deduplicated before scoring by replacing each formula instance with the unique identifier for the corresponding visually distinct formula and then removing duplicates using a greedy process starting at the top of the ranked list.  For example, if the single-symbol formula `$x$' were retrieved at ranks 1, 3, and 10, only the instance at rank 1 would be present after deduplication. To maximize comparability for later runs that might produce highly ranked documents that had not been judged, unjudged visually distinct formulae were removed from all runs before computing these measures, as proposed by Sakai \cite{sakai2007alternatives}.  Following Sakai's `prime' notation for this, the actual ARQMath measures are nDCG$'$, MAP$'$ and P$'$@10.  When computing MAP$'$ and P$'$@10, relevance grades N and P- were binarized as non-relevant and grades P+ and R were binarized as relevant.

Uniquely among these five test collections, ARQMath has a defined train/test split; for the other collections, results were reported at the original evaluation venue on the full collection. The ARQMath train sets were created after results had been reported, and are intended for use in training future systems.

\section{Effects of Complexity, Context, and Clustering}

We now use ARQMath-1/-2 and NTCIR-12 WFB to illustrate some consequences of these evaluation design choices. Table~\ref{tab:arq_ntcir_result} compares nDCG$'$ results for three systems on the 45 ARQMath-1 and 58 ARQMath-2 test queries, to nDCG$'$ results for the same systems using the 20 concrete NTCIR-12 WFB task queries. All three systems use the same base representation, but Tangent-S performs similarity computations on sub-trees, Tangent-CFT \cite{mansouri2019tangent} does so using formula embeddings (dense vectors), and Tangent-CFTED \cite{mansouri2020dprl} reranks Tangent-CFT results using tree edit distances between the full query and candidate SLTs and OPTs. Note that all systems achieved better nDCG$'$ in the NTCIR-12 collection. Here we explore possible reasons.   

\begin{table}[t!]
\caption{\label{tab:topic_correlation} Kendall's $\tau$ correlation by query complexity for nDCG$'$ system rankings.}
\scriptsize
\centering
 \resizebox{1.0\columnwidth}{!}{
\begin{tabular}{l | c | c | c | c   c  c}
\toprule
\sc Collection~~ & Teams & Runs & Mean Gap & \sc Low:Medium~~ & \sc Low:High~~ & \sc Medium:High\\
\midrule
\sc {ARQMath-1} & 4 & 10 & 0.053 & 0.80 & 0.77 & 0.76 \\ 
\sc {ARQMath-2} & 7 & 18 & 0.029 & 0.64 & 0.59 & 0.37 \\ 
\bottomrule
\end{tabular}
}
\end{table}

\begin{table}[!t]
\caption{\label{tab:arq_ntcir_result} NDCG$'$ for ARQMath and for NTCIR-12 WFB with undergrad student (UG), graduate student (Grad) or Aggregate judgments.}
\begin{center}
\begin{tabular}{l  | c c | c  c  c |  c c | c  c c}
\toprule
  & \multicolumn{5}{c}{\sc All instances} & \multicolumn{5}{c}{\sc Visually-distinct} \\
\cline{2-6}\cline{7-11}
  &\multicolumn{2}{c }{\sc ARQMath} & \multicolumn{3}{c}{\sc NTCIR-12 WFB} &
  \multicolumn{2}{c }{\sc ARQMath} & \multicolumn{3}{c}{\sc NTCIR-12 WFB} \\
\hline
Tangent &  -2 &  -1 &  Aggregate &  UG &  Grad & -2&  -1  &  Aggregate &  UG &  Grad\\
\hline
{-S}     & 0.49 & \textbf{0.69} & 0.78 & 0.70 & 0.75 & 0.49 &\textbf{0.69} & 0.79	&0.76	& 0.81\\
{-CFTED} & \textbf{0.57} &  0.56 & \textbf{0.89} & \textbf{0.80} &	\textbf{0.86}& \textbf{0.58} & 0.65 & \textbf{0.90} &	 \textbf{0.85} & \textbf{0.90}\\
{-CFT} & 0.56 & 0.53  & 0.86	& 0.79 &	0.81&	0.57&0.61&0.87	&	0.84&0.85\\
\bottomrule
\end{tabular}
\end{center}
\end{table}

{\bf Query Complexity.} 
Table \ref{tab:topic_correlation} illustrates the extent to which queries of different complexity result in ranking the same systems differently.  We compute Kendall's $\tau$ over all submitted runs, including the one baseline, and we count the baseline as a separate team.  Little effect from complexity is evident for ARQMath-1, but the rightmost two columns show that high-complexity queries result in different system rankings for ARQMath-2.  As the average difference in nDCG$'$ between adjacent runs when ranked by that measure (Mean Gap) indicates, smaller differences suffice to swap two systems in ARQMath-2, making it the better test collection for this analysis.  The average number of ARQMath-2 swaps between Medium:High queries is 4.3, almost twice the average of 2.6 swaps for Low:High topics. These results comport with prior analysis;
Mansouri et al.~\cite{mansouri2021ltr} have shown that different matching approaches (e.g., sub-tree, full-tree or embedding similarity) are better adapted to different complexity levels.  

Across different categories the top runs change in their rank position, and in some cases substantially. For high-complexity queries, the top run's nDCG$'$ is 0.09 higher than the rank 2 run, but is then 
ranked 9th for the low and medium complexity queries (respectively 0.07 and 0.06 lower than the top run). 
For medium-complexity queries, the top run
is ranked 10th and 5th for high and low-complexity queries, respectively. More generally, ARQMath-2 system rankings differed substantially by query complexity, and the top systems were different for each complexity level.

In addition to query complexities affecting rankings, complexity distributions differ in our test collections. Table~\ref{tab:complexities} shows 43\% of ARQMath queries (62/144) are low complexity, compared to just 20\% (4/20) for NTCIR-12 WFB queries. In ARQMath-2, more attention was paid to balancing query complexities.

\textbf{Context.}  In ARQMath-2, assessors were asked whether the retrieved formula was useful when interpreted in the post where the formula appeared;
{\bf even an exact match could be non-relevant} by that definition, whereas for ARQMath-1 and NTCIR-12, assessors were told that identical formulas were always relevant. For example, for topic B.289 in ARQMath-2, the original question specified that the formula query $x^n+y^n+z^n$ 
could take any real values for x, y, and z. Four of the pooled formula instances had appeared in contexts that limited the values to integers, and thus were judged as Non-Relevant. The one instance that had no limits on the values appeared in a context in which the formula was used for a purpose unrelated to the original question, and thus was also judged as Non-Relevant. As shown in Table \ref{tab:arq_ntcir_result}, with visually-distinct formulae, the nDCG$'$ value drops noticeably moving from NTCIR-12 WFB to ARQMath collections. Furthermore, all three systems have lower nDCG$'$ in ARQMath-2. This may be caused by Tangent systems ignoring surrounding context for formulae: their first retrieved formulae were deemed non-relevant for 9 (-CFT), 7 (-CFTED), and 20 (-S) topics. Notably, many of these non-relevant top hits had high visual similarity with the query (4/9 -CFT, 3/7 -CFTED, and 9/20 -S).

\begin{figure*}[!b]
\centering
{
\includegraphics[width=0.95\textwidth]{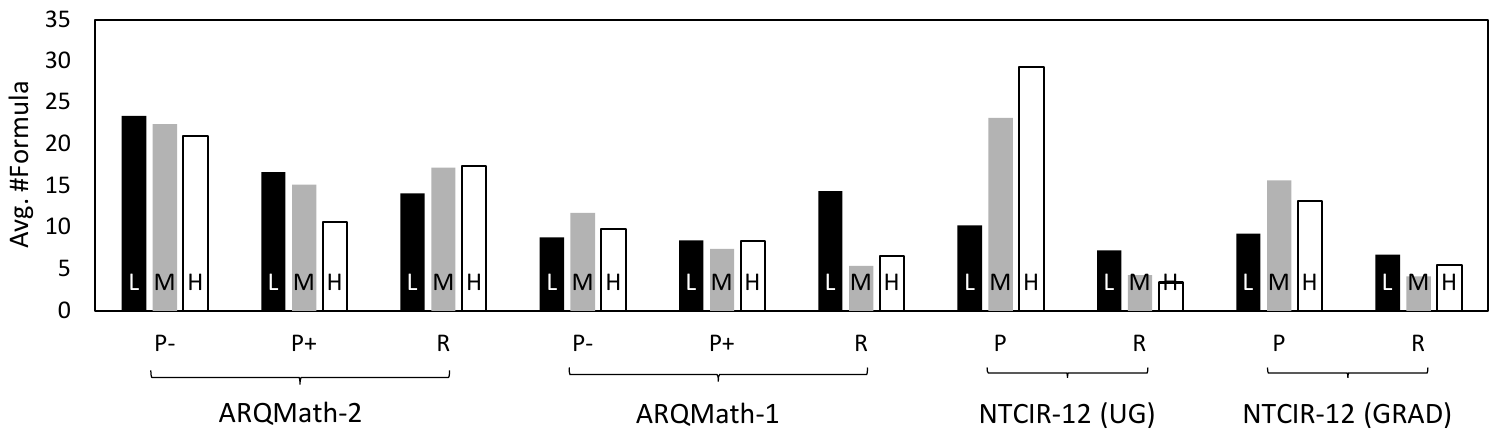}
}
\caption{Mean visually-distinct relevant formulae by query complexity (L/M/H).  UG/GRAD: undergrad/grad assessors. R indicates relevant, P partially relevant, P- \emph{low} partial relevance, and P+ \emph{high} partial relevance.}
\label{fig:complexity}
\end{figure*}

{\bf Clustering.} 
One effect of clustering formulae in ARQMath is increased visual diversity of formulas assessed for the ARQMath collections. 
As shown in Table~\ref{tab:annotated}, in ARQMath-1 and -2, the number of distinct formulae assessed on average was 134.9 and 139.8, respectively. This is noticeably higher than the average number of distinct formulae in the NTCIR-12 WFB test collection.

In Table~\ref{tab:arq_ntcir_result}, NTCIR-12 results are shown both for the original 3-level relevance grades (R, P, N) from each annotation team (undergraduate or graduate) and for the aggregated 5-level relevance grades (4, 3, 2, 1, 0) that were actually used at NTCIR-12. Results for evaluation using all formula instances (as reported for NTCIR) and for visually-distinct formula instances (as reported for ARQMath) are shown. For evaluation with visually distinct formulas for NTCIR-12, we performed max-aggregation for each of the three sets of relevance judgments in the same manner as was done at ARQMath. As can be seen, the NTCIR-12 and ARQMath-2 test collections prefer the Tangent-CFTED system, whereas the ARQMath-1 test collection prefers the Tangent-S system. These preferences hold true for both evaluation conditions (all instances, or visually distinct) and for all three sets of NTCIR-12 relevance judgments.

Table~\ref{tab:arq_result} summarizes results for ARQMath-1 and -2 for three Tangent family systems, Tangent-S, -CFTED, and -CFT. Here we consistently use the ARQMath test collections (the 45 ARQMath-1 and the 58 ARQMath-2 test queries) and the same evaluation measures; we scored either with all assessed formula instances, or with visually distinct formulae. It should be noted that Tangent-S is designed to retrieve exactly one instance for each visually distinct formula.

In ARQMath-1, P$'$@10 clearly prefers Tangent-CFTED when evaluating using all judged formula instances, but the same measure does not establish a clear preference between Tangent-S and Tangent-CFTED when evaluated using visually distinct formulae. It is the presence of visually identical duplicates among the top-10 that cause that difference; 10.5\% of visually distinct ARQMath-1 formulae have at least 2 judged instances. With the exception of nDCG$'$ values for Tangent-S, all systems obtain higher effectiveness values in all measures when considering formulae instances.\footnote{Results using ids for visually-distinct vs. formula instances can lead to results reordering by the {\tt trec\_eval}, which breaks ties based on identifier value.}

In ARQMath-2, however, this effect is not seen. For all the three effectiveness measures, Tangent-CFTED is preferred over other systems. Also, the gap in effectiveness measures using all the formula instances and visually-distinct formulae is much lower for Tangent-CFT and -CFTED compared to ARQMath-1. This might be caused by the ARQMath organizers making formula cluster (visual) ids available before the runs were submitted, and systems used those ids to reduce redundancy: submitted runs returned up to 5 instances for each visually distinct formula in ARQMath-2. Therefore, we can conclude that if systems do not consider diversity of formulas in their results, effectiveness measures can be different using visually-distinct formulae compared to formulae instances.

\begin{table*}[!ptb]
\caption{\label{tab:arq_result} ARQMath-1 and -2 results for Tangent-S, -CFTED, and -CFT.}
\begin{center}

\begin{tabular}{l  c  c  c  c  c  c }
\toprule
  & \multicolumn{3}{c}{\sc All instances} & \multicolumn{3}{c}{\sc Visually-distinct} \\
RUNS & \sc ~nDCG$'$ & \sc MAP$'$ & \sc P$'$@10 & \sc ~nDCG$'$ & \sc MAP$'$ & \sc P$'$@10 \\
\hline
 \multicolumn{7}{c}{ \bf ARQMath-1}\\
 \hline
 {Tangent-S}   & \textbf{0.69} &	0.47&	0.48 & \textbf{0.69} & \textbf{0.45} & \textbf{0.45} \\
 {Tangent-CFTED} &  0.65& \textbf{0.49} & \textbf{0.56} & 0.56&	0.39&	0.44 \\
 {Tangent-CFT} &  0.61	&0.43&	0.46&	0.53&	0.33&	0.35 \\
 \midrule
 \multicolumn{7}{c}{ \bf ARQMath-2}\\
 \hline
 {Tangent-S}   & 0.49 & 0.29&0.49	 & 0.49 & 0.27 & 0.42 \\
 {Tangent-CFTED} & \textbf{0.57} & \textbf{0.37} & \textbf{0.58} &\textbf{0.58} &	\textbf{0.38}&	\textbf{0.55} \\
 {Tangent-CFT} & 0.56 	& 0.35&	0.54&0.57&	0.36&	 0.52\\
\bottomrule
\end{tabular}
\end{center}
\end{table*}

\section{Conclusion}

We have compared six formula retrieval test collections, which differ in document sources, query characteristics, relevance definitions, assessor characteristics, relevance grades, and evaluation protocols. We have shown that some of these differences can affect the preference order between systems.  
Relative to the NTCIR test collections, the ARQMath test collections have both more queries and more diverse queries. Formulae judged for relevance are also more diverse, because only a few samples are assessed when many visually identical formulae are found. We have also highlighted differences in the way that relevance was assessed, illustrating a progression toward higher fidelity modeling of specific retrieval tasks.  
In future work, we hope to apply these insights to help design test collections in which consequential choices of the types we have illustrated can be made more intentionally.

\section*{Acknowledgments}
This material is based upon work supported by the Alfred P. Sloan Foundation under Grant No. G-2017-9827 and the National Science Foundation (USA) under Grant No. IIS-1717997.

% \bibliographystyle{plain} % We choose the "plain" reference style
% \bibliography{references} % Entries are in the refs.bib file

\end{document}